\documentclass[letterpaper,twocolumn,aps,amsmath,groupedaddress,floatfix,longbibliography]{revtex4-2}
\usepackage{amsmath}
\usepackage{amssymb}
\usepackage{mathdots}
\usepackage{graphicx}
\usepackage{hyperref}

\makeatletter

\pdfpageheight\paperheight
\pdfpagewidth\paperwidth

\usepackage{color}\usepackage{verbatim}
\usepackage{amsfonts}
\usepackage{epsfig}
\usepackage{ulem}
\usepackage{braket}
\usepackage{bm}
\graphicspath{{Figures/}}

\makeatother

\begin{document}
\title{Quantum Metric Induced Critical Current Anomaly in Flat Band Josephson Junctions}
\author{Zhong C.F. Li$^{1}$}\thanks{These authors contributed equally to this work.}
\author{Yuxuan Deng$^{1}$} \thanks{These authors contributed equally to this work.}
\author{Dmitri K. Efetov$^{2}$} \thanks{dmitri.efetov@lmu.de}
\author{K. T. Law$^{1}$} \thanks{phlaw@ust.hk}
\affiliation{1. Department of Physics, Hong Kong University of Science and Technology, Clear Water Bay, Hong Kong, China}
\affiliation{2. Fakult\"at f\"ur Physik, Ludwig-Maximilians-Universit\"at, Schellingstrasse 4, M\"unchen 80799, Germany}

\begin{abstract}
In well-established theories of Josephson junctions, the superconducting critical current \( I_\mathrm{c} \) increases as the normal state conductance \( \mathcal{G} \) increases. However, in a recent experiment in twisted bilayer graphene (TBG) based Josephson junctions, unexpectedly, it was observed that the increase of the critical current is accompanied by a decrease of the normal state conductance. We call this phenomenon the critical current anomaly. In this work, we point out that in the TBG-based Josephson junction, due to the suppression of the conventional Josephson current by the flatness of the band and the quantum metric enabled Josephson current (QMJC), the critical current anomaly can occur.  The QMJC appears if the quantum metric length is comparable or longer than the junction length.  We show that both  \( \mathcal{G} \) and   \( I_\mathrm{c} \) have the conventional and the quantum metric contributions, and there are parameter regimes in which  \( I_\mathrm{c} \) increases even when  \( \mathcal{G} \) decreases. We first demonstrate the critical current anomaly by a simple modified Lieb-lattice model both analytically and numerically. The incredible consistency with the experimental results is demonstrated using a realistic six-band model of twisted bilayer graphene. Therefore, we suggest that the critical current anomaly observed in the experiment provide strong evidence of QMJC which were ignored in well-established theories of Josephson junctions.

\end{abstract}

\maketitle

\begin{figure}[ht]
    \centering
    \includegraphics[width=\linewidth]{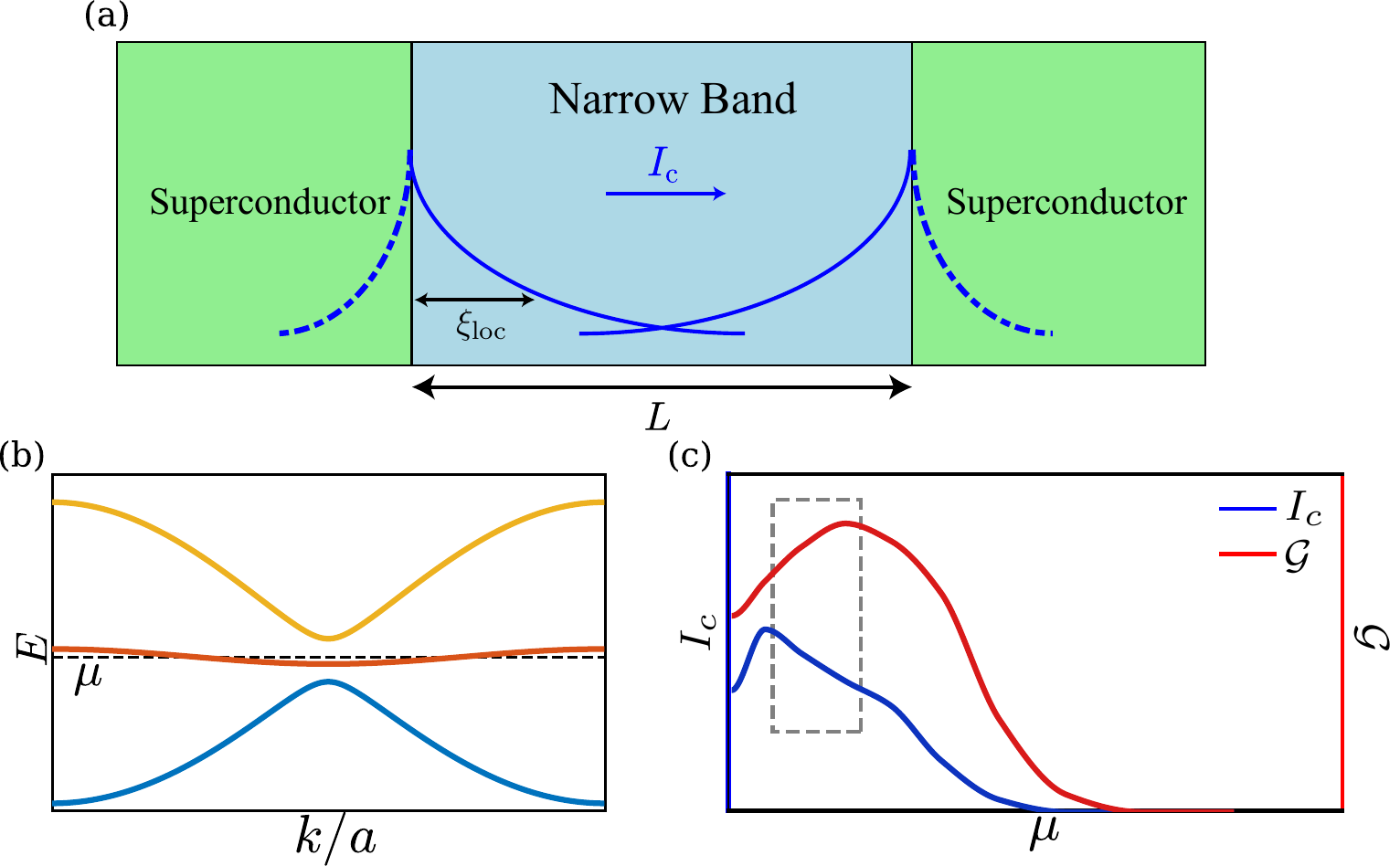}
    \caption{(a) The schematic figure of a flat band Josephson junction. The weak link is a narrow band material with nontrivial quantum metric. The leads are superconducting (SC) for calculating the Josephson current and metallic for calculating the normal-state conductance.  The blue lines at the lead/weak link interfaces represent the interface states with the decay length $\xi_{\mathrm{loc}}$ governed by the quantum metric length of the weak link. The interface states can carry Josephson current when the quantum metric length is long enough that the interface states are hybridized. (b) The schematic plot for the band structure of the weak link, where $\mu$ is the chemical potential and $k$ is the wave vector. (c) The schematic figure of the critical current anomaly highlighted by the dashed box, where $\mathcal{G}$ (red line) and $I_\mathrm{c}$ (blue line) are functions of chemical potential. The anomaly appears when $I_\mathrm{c}$ increases while $\mathcal{G}$ decreases as observed in the experiment \cite{ccb4-tqxq}. }
    \label{Fig1}
\end{figure}

The emergence of flat bands in engineered moiré materials has opened new avenues for exploring correlated and quantum geometrical phenomena \cite{bistritzer2011moire,cao2018unconventional,2018Natur.556...80C,2019Natur.574..653L,2019Sci...363.1059Y,PhysRevB.99.195455,2019SciA....5.9770C,2019PhRvX...9c1049H,balents2020superconductivity,PhysRevLett.124.076801,2020Natur.583..375S,2020NatMa..19.1265A,zondiner2020cascade,2020NatPh..16..926S,2021NatRM...6..201A,2021PhRvL.126b7002P,park2021tunable,2021PNAS..11806744V,rossi2021quantum,PhysRevLett.132.026002,PhysRevLett.130.216003,zhang2023visualizing,2023arXiv231015558D,crepel2024chiral,PhysRevLett.132.196202,PhysRevLett.132.036001,PhysRevX.15.021056,yu2025quantumgeometryquantummaterials,verma2025quantumgeometryrevisitingelectronic,park2025ferromagnetism,dong2025flatbandexcitonsthreedimensional}. In particular, the study of quantum metric effects in flat band superconductors has been a very important topic in recent years \cite{peotta2015superfluidity,PhysRevLett.117.045303,torma2018quantum,PhysRevB.98.134513,2019PhRvL.123w7002H,xie2020topology,PhysRevB.101.060505,PhysRevB.103.144519,verma2021optical,PhysRevB.106.014518,2022PhRvL.128h7002H,PhysRevB.106.184507,torma2022superconductivity,2022PhRvL.128h7002H,PhysRevLett.131.240001,jiang2025superfluid,penttila2025flat}. Recently,  the breakdown of the well-established relation between the normal-state conductance ($\mathcal{G}$) and the critical Josephson current ($I_\mathrm{c}$) in magic angle twisted bilayer graphene (TBG) Josephson junctions was experimentally observed \cite{ccb4-tqxq}. By electrostatically tuning the chemical potential of the TBG, it was found that $\mathcal{G}$ can be significantly suppressed, while $I_\mathrm{c}$ is simultaneously enhanced. The schematic illustration of this critical current anomaly and the theoretical calculations are shown in Fig.~\ref{Fig1}(c) and Fig.~\ref{Fig5}(b) respectively. This anomalous relation between $\mathcal{G}$  and $I_\mathrm{c}$ stands in sharp contrast to conventional theories, which expect that the decreases in the normal state conductance $\mathcal{G}$ always result in the decreases of the critical Josephson current $I_\mathrm{c}$ \cite{PhysRevLett.66.3056,Beenakker_JJ,Datta_1995,tafuri2019fundamentals}. For example, at zero temperature, $I_\mathrm{c}=\frac{\pi\Delta_\mathrm{sc}}{e}\mathcal{G}$ in the ballistic quantum point contact regime, while $I_\mathrm{c}=\frac{\pi\Delta_\mathrm{sc}}{2e}\mathcal{G}$ in the diffusive tunnel junction regime \cite{Beenakker_JJ,tafuri2019fundamentals}, where $\Delta_\mathrm{sc}$ is the superconducting pairing gap of the superconducting leads. In this work, we point out the quantum metric origin of the critical current anomaly. 

In the following sections, we explain that the anomalous relation between $\mathcal{G}$ and $I_\mathrm{c}$ arises from the coexistence of two distinct transport mechanisms in magic angle TBG. The first is the conventional mechanism, where the Josephson current is mediated by the propagating modes of the weak link related to the band dispersion. The second is a new mechanism with quantum metric origin,  where the Josephson current is mediated by interface states which appear at the superconductor/weak link interfaces as illustrated in Fig. \ref{Fig1}(a), where the decay length of the interface states is governed by the quantum metric length $\xi_\mathrm{QM} $ defined in Eq.\ref{QML} \cite{PhysRevResearch.7.023273,l3c7-knqm, supp}. The wavefunctions of the interface states are shown in Fig.2 of the Supplemental Material. Since the interface state mediated Josephson current goes to zero without quantum metric, we call this quantum metric enabled Josephson current (QMJC). 



For the conventional contribution due to the  band dispersion, $\mathcal{G}$ is non-decaying in the ballistic regime \cite{de2018superconductivity}, while the band dispersion contribution to the critical Josephson current $I_\mathrm{c}$ decays exponentially with junction length as $e^{-L/\xi_{v_\mathrm{F}}}$.  Here, $\xi_{v_\mathrm{F}} = \hbar v_\mathrm{F}/(2\pi k_\mathrm{B} T)$ is the coherence length associated with the Fermi velocity $v_\mathrm{F}$ as shown by Bardeen and Johnson more than five decades ago \cite{PhysRevB.5.72}. However, the effect of quantum metric \cite{provost1980riemannian} was ignored in well-established theories of Josephson junctions \cite{PhysRevB.5.72,tafuri2019fundamentals}. 


On the other hand, the quantum metric effects come in through the interface states localized at the lead/weak link interfaces [See Fig.1(a)]. The interface states contribute to both $\mathcal{G}$ and $I_\mathrm{c}$. Importantly, the interface state contribution to $\mathcal{G}$ and $I_\mathrm{c}$ decay exponentially as $e^{-L/\xi_\mathrm{loc}}$, where $\xi_\mathrm{loc} = 8\xi_\mathrm{QM} $ is the localization length of the interface states \cite{PhysRevResearch.7.023273} and independent of $v_\mathrm{F}$ and temperature. As shown in the Supplemental Material \cite{supp}, $\xi_\mathrm{loc}$ is proportional to the quantum metric length \cite{hu2025anomalous,PhysRevResearch.7.023273,l3c7-knqm} which is given explicitly in Eq.~\ref{QML}.  In the regime where $\xi_{v_\mathrm{F}} \ll \xi_\mathrm{loc} \lesssim L$, which is achievable in TBG junctions~\cite{ccb4-tqxq}, $\mathcal{G}$ is dominated by the band-dispersion contribution, while $I_\mathrm{c}$ is dominated by the quantum metric enabled component. In the experiment, by gating $\mu$ towards the charge neutrality point,  $\mathcal{G}$ is suppressed, whereas $I_\mathrm{c}$ is enhanced since the dominant QMJC peaks near charge neutrality. This interplay between the conventional and QMJC results in the observed critical current anomaly as illustrated in Fig.~\ref{Fig5}(b), which is obtained by using a realistic six-band tight-binding model for the calculations which matches the experimental results well without fine tuning of parameters. Therefore, we suggest that the critical current anomaly observed in the experiment provide strong evidence of QMJC.

\begin{figure}
    \centering
    \includegraphics[width=\linewidth]{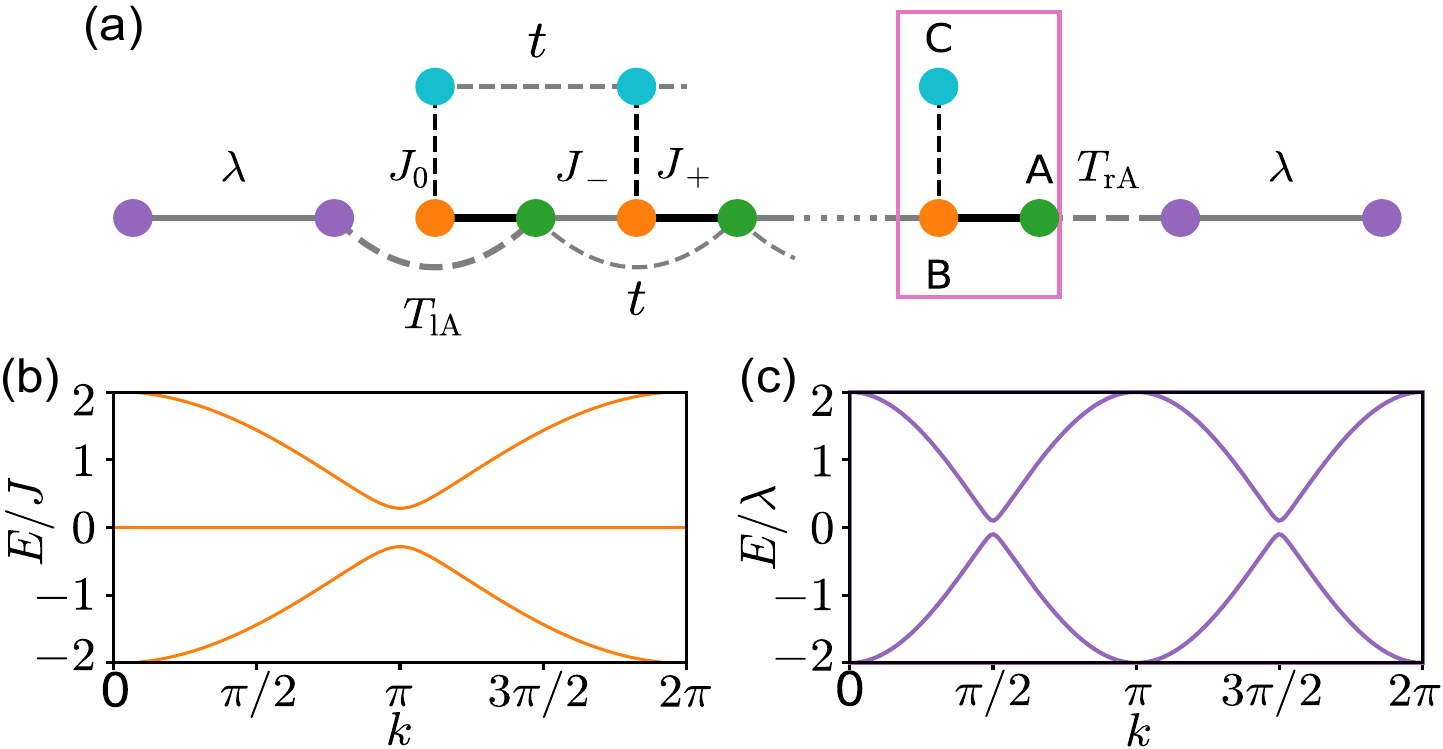}
    \caption{(a) The lattice structure of the Josephson junction. (b) The band structure of Lieb lattice. For visualization, $J=10\Delta_\mathrm{sc}$ and $\delta=0.1$. (c) The quasiparticle spectrum of the superconducting leads, which is $\epsilon(k)=\pm\sqrt{(2\lambda\cos{ka})^2+\Delta_\mathrm{sc}^2}$. For visualization, $\lambda=10\Delta_\mathrm{sc}$. }
    \label{Fig2}
\end{figure}

\begin{figure}[t]
    \centering
    \includegraphics[width=\linewidth]{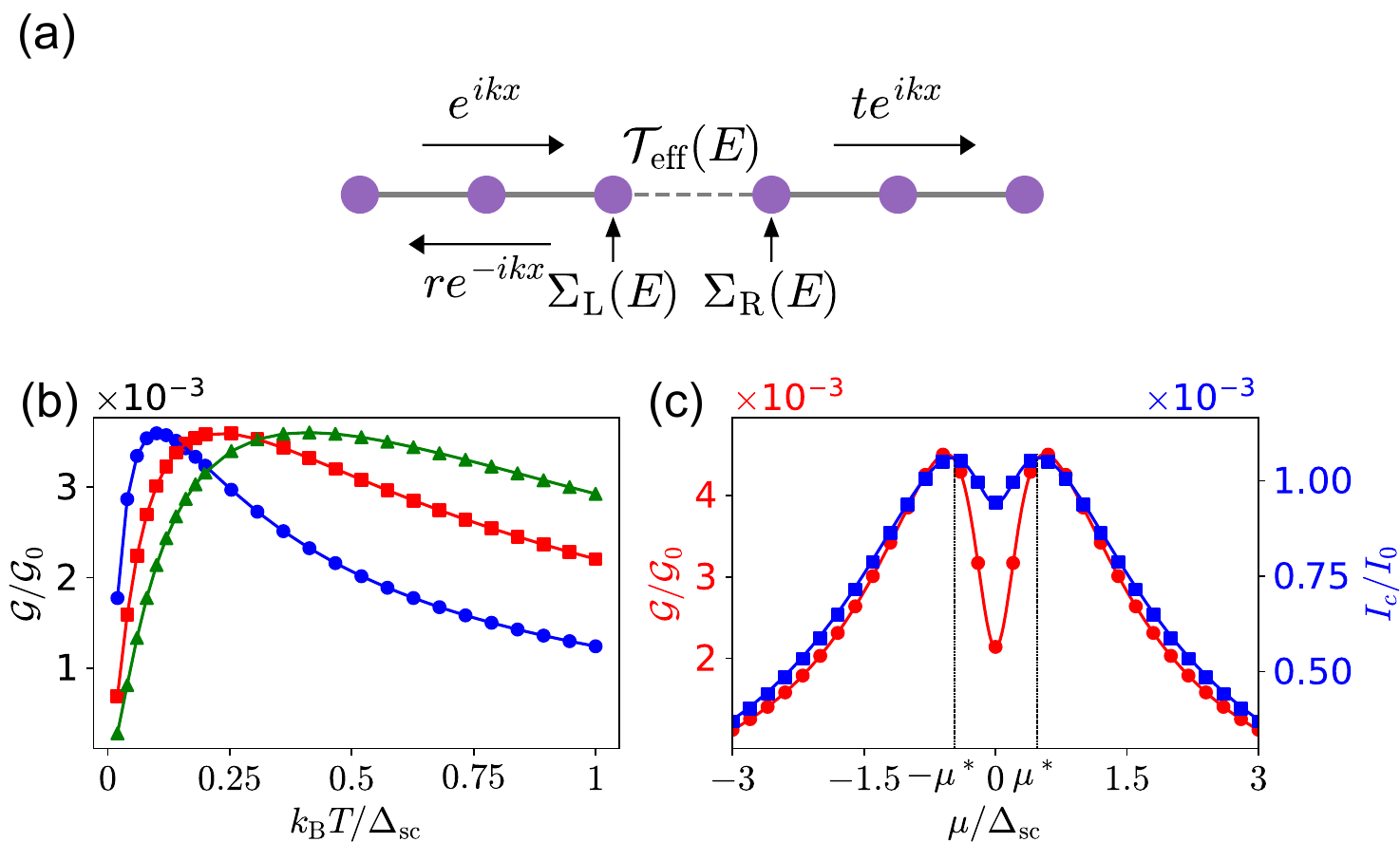}
    \caption{(a) A schematic figure of the effective models calculating transmission rate and critical Josephson current, respectively. (b) Temperature-dependence of the normal-state conductance $\mathcal{G}$, where $\mathcal{G}_0=e^2/h$ is the conductance unit. For comparison, the superconducting pairing gap $\Delta_\mathrm{sc}$ at zero temperature of the superconducting leads is used as energy scale, even the leads used for calculating $\mathcal{G}$ are metallic. The coupling strengths $T_\mathrm{A}$ for the blue line, red line and green line are $50\Delta_\mathrm{sc}$, $75\Delta_\mathrm{sc}$ and $100\Delta_\mathrm{sc}$, respectively. The analytical results (solid lines) and the numerical results (discrete data points) match each other perfectly. (c) The comparison between conductance and critical Josephson current. The numerical results of $I_\mathrm{c}$ and $\mathcal{G}$ are represented by the blue squares and red circles, respectively. The blue line and red line are the analytical results of $I_\mathrm{c}$ and $\mathcal{G}$. In (b) and (c), $\delta=0.02$, the band gap $2\sqrt{2}J\delta$ of Lieb lattice is chosen to be much larger than $\Delta_\mathrm{sc}$ and $\lambda$, and the hopping amplitudes in metallic and superconducting leads are the same $\lambda=100\Delta_\mathrm{sc}$. In (c), the coupling strength $T_\mathrm{A}=100\Delta_\mathrm{sc}$ }
    \label{Fig3}
\end{figure}

In the rest of this paper, we first demonstrate the properties of the QMJC which is mediated by the interface states in a 1D Lieb lattice model with exact flat bands.  The exact flat band model allows us to calculate and understand the QMJC analytically which matches the numerical calculations perfectly as shown in Fig.~\ref{Fig3}(c). We then introduce a finite band dispersion and demonstrate the critical current anomaly in the 1D Lieb lattice model. Finally, we show that the critical current anomaly can also be observed in two-dimensional Lieb lattice as well as the more realistic six-band model of twisted bilayer graphene \cite{PhysRevB.99.195455}.
\\
\\
\textbf{Results}\\
\textbf{Flat band Josephson junction model with 1D Lieb Lattice}. In this section, we introduce a one-dimensional Lieb lattice model with a flat band and nontrivial quantum metric to describe the weak link. One reason to start with a Lieb lattice is that TBG can indeed be described by a six-band lattice as shown in Ref.\cite{PhysRevB.99.195455}. Moreover, this simple lattice model allows us to calculate  $\mathcal{G}$ and $I_\mathrm{c}$ both numerically as well as fully analytically for illustrating the QMJC.  As shown in Fig.~\ref{Fig2}(a), next-nearest-neighbor hopping $t$ on sublattices A and C is introduced so that the flat band acquires a finite dispersion as in the case of TBG. The demonstration of the critical current anomaly using a two-dimensional Lieb lattice is in the Supplemental Material \cite{supp}. Demonstration of the critical current anomaly using a six-band realistic Hamiltonian for TBG is shown in a later section.

Coming back to the one-dimensional Lieb lattice, we define the creation and annihilation operators as $a^\dagger_{n\alpha\sigma}$ and $a_{n\alpha\sigma}$ for electrons with spin $\sigma = \uparrow/\downarrow$ on orbital $\alpha = \mathrm{A}, \mathrm{B}, \mathrm{C}$ in unit cell $n$. The Hamiltonian takes the form:

\begin{equation}\label{H_Lieb}
\begin{aligned}
    H_\mathrm{Lieb}=&\sum^N_{n=0}\sum_\sigma(J_+a^\dagger_{n\mathrm{B}\sigma}a_{n\mathrm{A}\sigma}+J_0a^\dagger_{n\mathrm{B}\sigma}a_{n\mathrm{C}\sigma}\\
    &+J_{-}a^\dagger_{n+1,\mathrm{B}\sigma}a_{n\mathrm{A}\sigma}+ \mathrm{h.c.})\\
    &+\sum_{n=0}^{N}t(a^\dagger_{n+1,\mathrm{A}\sigma}a_{n\mathrm{A}\sigma}+a^\dagger_{n+1,\mathrm{C}\sigma}a_{n\mathrm{C}\sigma}+\mathrm{h.c.}),
\end{aligned}
\end{equation}
where $J_\pm=(1\pm\delta)J$, $J_0=2i\delta J$, $0<\delta<1$, and the next-nearest-neighbor hopping amplitude $t$ is taken to be real. Here, the length of the weak link is $L=Na$, where $a$ is the lattice constant. If $t=0$, there is an exactly flat band in the Lieb lattice with band gap $2\sqrt{2}\delta J$, as shown in Fig.~\ref{Fig2}(b). For nonzero $t$, the originally flat band $\bar{n}$ acquires dispersion $\epsilon_{\bar{n}k}=2t\cos{k}$. For the 1D Lieb lattice, the real part of the quantum metric tensor associated with the flat band can be written as $g^{0}(k)$ such that 
\begin{equation}\label{quantum_metric}
    g^{0}(k)=\mathrm{Re}\langle\partial_{k}u_{0}(k)|(1-\vert u_{0}(k)\rangle\langle u_{0}(k)\vert)|\partial_{k}u_{0}(k)\rangle,
\end{equation}
where $|u_{0}(k)\rangle$ is a Bloch state of the Hamiltonian. With $g^{0}(k)$, the quantum metric length of the Lieb lattice is defined as:
\begin{equation}\label{QML}
\xi_\mathrm{QM} = \int_{-\pi/a}^{\pi/a} g^{0}(k)\frac{dk}{2\pi} = \frac{a}{16\sqrt{2}\delta}.
\end{equation}

As shown in Ref.\cite{PhysRevResearch.7.023273}, and the Supplemental Material (Section II) \cite{supp}, the localization length  of the interface state $\xi_{\mathrm{loc}}$ equals to $8\xi_\mathrm{QM}$. To calculate the normal state conductance, the Lieb lattice is coupled to two one-dimensional metallic leads through A orbitals at the leftmost and rightmost unit cells with coupling strength $T_\mathrm{lA}$ and $T_\mathrm{rA}$, respectively, as shown in Fig.~\ref{Fig2}(a). The band dispersion of the metallic leads is given by $2\lambda\cos{k}$. For the calculation of the Josephson current, the superconducting order parameters $\Delta_\mathrm{l/r}=\Delta_{\mathrm{sc}}e^{i\phi_{\mathrm{l/r}}}$ are introduced for the left and right leads respectively, where $\phi=\phi_\mathrm{l}-\phi_\mathrm{r}$ is the phase difference between the two superconducting leads, and $\Delta_\mathrm{sc}$ represents the superconducting pairing gap. The band dispersion of the superconducting leads is shown in Fig.~\ref{Fig2}(c). Using $c^\dagger_{i\sigma}$ and $c_{i\sigma}$ to denote the creation and annihilation operators for electrons with spin $\sigma$ at site $i$ in the lead, the Hamiltonian $H_\mathrm{sc}$ of the superconducting leads are given by:

\begin{equation}\label{H_leads}
    \begin{aligned}
        H_{\mathrm{sc,L/R}}= &\sum_{\langle ij\rangle,\sigma}\lambda (c_{i\sigma}^{\dagger}c^{}_{j\sigma}-c_{i\sigma}c^{\dagger}_{j\sigma}+\mathrm{h.c.})\\
        & +\!\sum_{i}(\Delta_{\mathrm{l/r}}c_{i\uparrow}^{\dagger}c_{i\downarrow}^{\dagger}+\mathrm{h.c.}),
    \end{aligned}
\end{equation}
where $\langle ij\rangle$ indicates that $i$ and $j$ are neighboring sites. In the following discussion, the band gap of the Lieb lattice, $2\sqrt{2}\delta J$, is chosen to be much larger than $t$ and $\Delta_\mathrm{sc}$ . The Hamiltonian of the metallic leads is obtained by setting the superconducting pairing potential to be zero. For the one-dimensional model, the chemical potential of the leads is chosen at the middle of the band, which is zero in Eq.~(\ref{H_leads}). The coupling terms between the weak link and the leads are described by
\begin{equation}
H_{\mathrm{couple}}=\sum_\sigma(T_\mathrm{A}a^\dagger_{0A\sigma}c_{-1\sigma}+T_\mathrm{A}a^\dagger_{NA\sigma}c_{N+1\sigma}+\mathrm{h.c.})
\end{equation}
where the sites $i=-1$ and $i=N+1$ correspond to the right end of the left lead and the left end of the right lead, respectively. The coupling amplitude $T_\mathrm{A}$ can be comparable to $\lambda$ and we assume $\lambda\gg\Delta_\mathrm{sc}$ in the following calculations. 

\textbf{Quantum metric enabled Josephson current}. In this section, we study the quantum metric contributions to $\mathcal{G}$ and $I_\mathrm{c}$ in a flat band junction, using the one-dimensional Lieb lattice described by Eq.~(\ref{H_Lieb}). We start with the exact flat band limit with $t=0$. This simple model allows us to calculate the normal state conductance as well as the Josephson current fully analytically. The analytical results match the numerical results perfectly as shown in Fig.~\ref{Fig3}(b) and Fig.~\ref{Fig3}(c).  It is important to note that with an exact flat band, there is no conventional contribution to the Josephson current as the Fermi velocity is zero. We can clearly see below that the Josephson current is governed by the quantum metric length $\xi_\mathrm{QM}$ in the analytical results. With small quantum metric length compared to the junction length, the Josephson current vanishes. Therefore, we call this Josephson current originated from the coupling of the interface states \cite{PhysRevResearch.7.023273, supp}, the quantum metric enabled Josephson current (QMJC).

To calculate the conductance, we consider an incident wave with wave vector $k$ and energy $E=2\lambda\cos k$ from the left metallic lead. The transmitted wave in the right metallic lead is written as $t(E)e^{ikx}$, where $t(E)$ is the transmission amplitude and $T(E)=|t(E)|^2$ is the transmission rate. Once $T(E)$ is obtained, the conductance is given by
$\mathcal{G}=\mathcal{G}_0\frac{1}{k_\mathrm{B}T}\int_{-\infty}^{+\infty}T(E)f(E)\left[1-f(E)\right]dE$ \cite{girvin2019modern},
where $\mathcal{G}_0=e^2/h$ is the conductance quantum and $f(E)$ is the Fermi-Dirac distribution at temperature $T$.

Using the lattice model established above as an example, the localization length
\begin{equation}
    \xi_{\mathrm{loc}}=a/(2\sqrt{2}\delta)=8\xi_{\mathrm{QM}},
\end{equation}
characterizes the spatial extent of the interface states, where the factor 8 is model dependent \cite{PhysRevResearch.7.023273}. By integrating out the degrees of freedom in the weak link, we obtain an effective description involving only the surface sites of the two leads, as illustrated in Fig.~\ref{Fig3}(a) \cite{supp}. In this effective Hamiltonian, the weak link induces an effective coupling $\mathcal{T}_{\mathrm{LR}}(E)$ between the left and right interfaces, and the two leads introduce onsite self-energy corrections $\Sigma_{\mathrm{L/R}}(E)$ at the left and right interfaces. The transmission rate for each spin channel can be written as \cite{supp}
\begin{equation}\label{T(E)}
    T(E)=\frac{4\lambda^2|\mathcal{T}_\mathrm{LR}|^2}{\big(\lambda^2+|\mathcal{T}_\mathrm{LR}|^2-\Sigma_\mathrm{L}\Sigma_\mathrm{R}\big)^2+\lambda^2\big(\Sigma_\mathrm{L}+\Sigma_\mathrm{R}\big)^2},
\end{equation}
where $|\mathcal{T}_\mathrm{LR}| \propto |T_\mathrm{A}|^2 e^{-L/\xi_\mathrm{loc}}$. The details of calculation is provided in the Supplemental Material \cite{supp}. 

Since the band is exactly flat, the conventional dispersive contribution to transmission vanishes. The transmission therefore originates from the overlap of interface states, whose localization length is controlled by the quantum metric length \cite{PhysRevResearch.7.023273}. We refer to the resulting contribution to the conductance as the quantum metric contribution. In the low-temperature regime, increasing temperature can enhance the quantum metric contribution to the conductance $\mathcal{G}$ by increasing the occupation of the interface-state transport channel, whereas the dispersive contribution is suppressed. This temperature dependence of $\mathcal{G}$ is shown in Fig.~\ref{Fig3}(b). According to Eq.~(\ref{T(E)}), the transmission probability $T(E)$ reaches its maximum at $\mu^* = \sqrt{\Sigma_\mathrm{L}\Sigma_\mathrm{R}} E/\lambda$. Consequently, the quantum metric contribution as a function of the chemical potential $\mu$ of the weak link has a maximum at $|\mu^*|\neq 0$, namely away from the band center, as shown in Fig.~\ref{Fig3}(c). It is important to note that the analytical result of Eq.(\ref{T(E)}) matches the numerical results obtained by the lattice Green's function ( See Supplemental Material Section III) approach extremely well. 

The QMJC can also be calculated from the free energy of the weak link. The Josephson current as a function of the phase difference is related to the phase-dependent part of the free energy by $I(\phi)=\frac{2e}{\hbar}\frac{\partial F(\phi)}{\partial\phi}$. The free energy of the Josephson junction can be calculated using the effective model shown in Fig.~\ref{Fig3}(a), where the degrees of freedom in the superconducting leads except for the surface sites are integrated out \cite{supp}, giving rise to an effective two-site model. Spin degrees of freedom are degenerate, and thus are not written explicitly. The superconducting critical current obtained is
\begin{equation}
\label{Ic}
    I_\mathrm{c}=\frac{2e}{\hbar}k_\mathrm{B}T\frac{|T_\mathrm{LR}|^2}{\lambda^2}\sum^{+\infty}_{n=0}\frac{4\Delta_\mathrm{sc}^2}{\omega_n^2+\Delta_\mathrm{sc}^2}f_n(\mu),
\end{equation}
where $\omega_n=(2n+1)\pi k_{\mathrm{B}}T (n=1,2, \cdots)$ are the $n^{\mathrm{th}}$ Matsubara frequency, and $f_n(\mu)$ is a coefficient depending on the self-energy of the leads and the Matsubara frequency. The full expression for $f_n(\mu)$ and $I_c$ is provided in the Supplemental Material \cite{supp}. Approximately, as a function of the chemical potential $\mu$, both the critical current and the conductance reach their maximum values at $|\mu^*|$, as shown in Fig.~\ref{Fig3}(c). 

These results demonstrate that the quantum metric contributions to the Josephson current and conductance exhibit qualitatively the same dependence on the chemical potential $\mu$, and both are suppressed by the exponential decay factor $e^{-2L/\xi_\mathrm{loc}}$, where $\xi_\mathrm{loc} = 8 \xi_\mathrm{QM}$.

\textbf{Critical Current Anomaly}. As shown above in Fig.~\ref{Fig3}(c), there is no critical current anomaly in the exact flat band limit. The anomaly appears once a small additional dispersion is introduced into the otherwise flat band, for example by taking $t\neq 0$ in Eq.~(\ref{H_Lieb}). The additional dispersion creates a conventional transport channel associated with the finite group velocity of the narrow band, which can be controlled by adjusting $t$, while the quantum-metric channel continues to originate from the interface-state contribution inherited from the flat-band geometry and is tunable via the parameter $\delta$. The anomaly follows from the fact that these two channels enter the conductance $\mathcal{G}$ and the critical current $I_\mathrm{c}$ in qualitatively different ways.

\begin{figure}
    \centering
    \includegraphics[width=\linewidth]{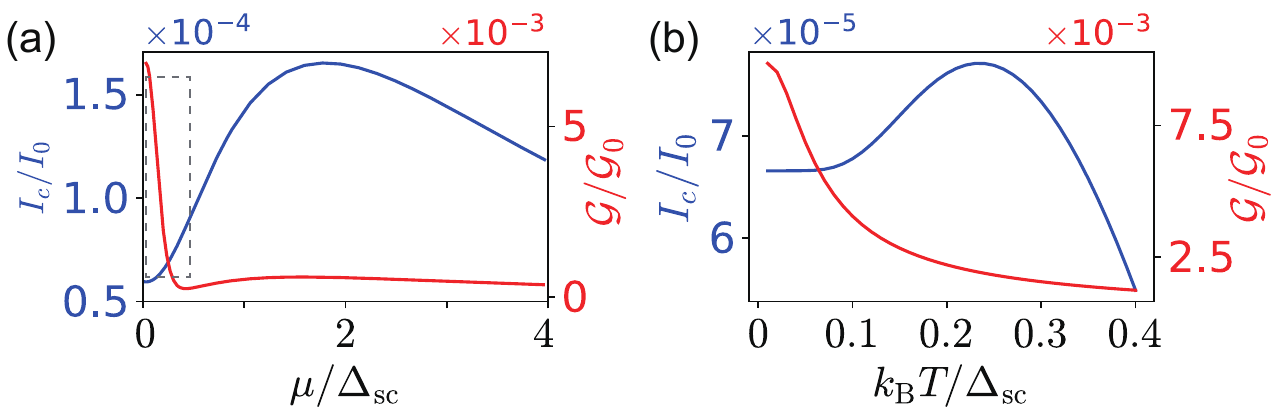}
    \caption{(a) The critical current anomaly probed by tuning the chemical potential $\mu$ at $k_\mathrm{B}T=0.05\Delta_\mathrm{sc}$. (b) The critical current anomaly probed by tuning the temperature $T$ at $\mu=0$. For critical-current calculations at different temperatures, the superconducting pairing gap depends on temperature as $\Delta_\mathrm{sc}(T)=\Delta_\mathrm{sc}\tanh(1.74\sqrt{T_\mathrm{c}/T-1})$, where $T_\mathrm{c}$ is the critical temperature of the superconducting leads. For the Lieb lattice, $\delta=0.05$, and the band gap $2\sqrt{2}J\delta$ is chosen to be the largest energy scale so that the narrow band is isolated. In (a) and (b), $\lambda=100\Delta_\mathrm{sc}$, $T_\mathrm{A}=100\Delta_\mathrm{sc}$, $t=0.05\Delta_\mathrm{sc}$, and $L=20a$.}
    \label{Fig4}
\end{figure}

According to the effective model shown in Fig.~\ref{Fig3}(a), the transmission rate $T(E)\propto|\mathcal{T}_\mathrm{LR}(E)|^2$ \cite{supp}. With a finite dispersion added to the otherwise flat band, the normal-state conductance consists of two contributions: the conventional contribution from the band dispersion and the quantum-metric contribution from the interface states. The conventional contribution is carried by the dispersive states of the weak link and is therefore not exponentially suppressed with junction length. In contrast, the quantum metric contribution is mediated by interface states and reaches maximum at $|\mu|=\mu^*$. However, the quantum metric contribution retains the real-space suppression factor $e^{-2L/\xi_\mathrm{loc}}$. Consequently, even a weak additional dispersion can strongly modify the conductance in Fig.~\ref{Fig4}(a): once the dispersive channel is available, $\mathcal{G}$ is mainly dominated by the dispersive contribution and varies rapidly as $\mu$ moves away from the band center.


The critical current is affected by the dispersion and the quantum metric in a different way. Both contributions are subject to exponential suppression in real space. The conventional contribution is characterized by the coherence length $\xi_{v_{\mathrm{F}}}=\hbar v_\mathrm{F}/2\pi k_\mathrm{B}T$, whereas the quantum metric contribution is controlled by the much longer scale $\xi_\mathrm{loc} = a/(2\sqrt{2}\delta)\gg\xi_{v_{\mathrm{F}}}$. In the regime considered here, the supercurrent is therefore limited by the factor $e^{-L/\xi_\mathrm{loc}}$, so the additional dispersion changes $I_\mathrm{c}$ only slightly, even though the dispersion strongly modifies $\mathcal{G}$.

This difference between the conductance and the critical current explains the critical current anomaly. The conductance is highly sensitive to the dispersive channels, whereas the critical current remains dominated by the QMJC because the conventional Josephson current decays rapidly across the junction. As a result, the normal-state conductance can vary strongly whereas the critical current changes only weakly, producing the behavior shown in Fig.~\ref{Fig4}(a). The detailed crossover analysis and the corresponding localization-length formulas are provided in the Supplemental Material \cite{supp}.

The same distinction between the conductance and the critical current explains the temperature-driven anomaly in Fig.~\ref{Fig4}(b). When the chemical potential is tuned near the middle of the narrow band, the conductance is mainly governed by the conventional dispersive channel and therefore decreases with increasing temperature \cite{tafuri2019fundamentals}. In contrast, the critical current remains controlled by the quantum metric as long as $\xi_\mathrm{loc}\gg\xi_{v_\mathrm{F}}$, so that $I_\mathrm{c}$ can increase in the low-temperature regime ($k_\mathrm{B}T\ll\Delta_\mathrm{sc}$) \cite{PhysRevResearch.7.023273}.

\begin{figure*}[t]
    \centering
    \includegraphics[width=\linewidth]{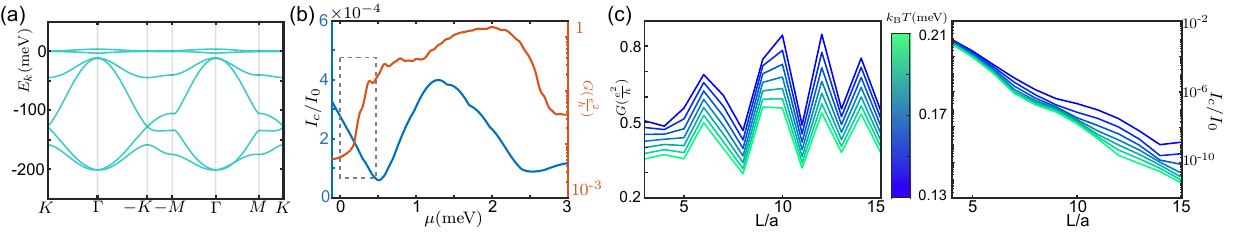}
    \caption{(a) Band structure of 6-band model of magic angle twisted bilayer graphene. (b) The critical current $I_\mathrm{c}$ and the resistance $R$ as a function of chemical potential $\mu$ in the six-band model. The parameters in (b) are set to be $\Delta_{\mathrm{sc}}=1$meV, $k_{\mathrm{B}}T = 0.15$meV, $\lambda = 10$ meV, $\mu_{\mathrm{Lead}}=0$. The coupling between the leads and the weak link is uniform from the lead to all orbitals in the weak link, and the amplitude is set to be $10$meV. The gap introduced between narrow bands is $1.9$meV. The size of the junction is $5\times30$ unit cells on $x$ and $y$ directions, respectively. (c) The conductance (left panel) and the critical current (right panel) against the junction length $L$. Conductance oscillates as the junction length increases, while the critical current decays exponentially as the junction length increases. }
    \label{Fig5}
\end{figure*}
\
\textbf{Realistic 2D calculation}. To show that the critical current anomaly is not a special feature of the quasi-one-dimensional toy model, we next study a realistic two-dimensional weak link based on the faithful six-band tight-binding model of magic-angle twisted bilayer graphene. In this calculation, the weak link is described by
\begin{equation}
H = H_{6\text{-band}} + \Delta H, \label{eq: 6-band TBG}
\end{equation}
where $H_{6\text{-band}}$ captures the realistic narrow-band structure of TBG and $\Delta H$ is a gap-opening perturbation that lifts the degeneracy between the two narrow bands near the $K$ points. We further consider a finite-width Josephson junction geometry with superconducting leads attached to the TBG weak link, so that both the normal-state conductance and the critical current can be evaluated in a realistic 2D model. As shown in Fig.~\ref{Fig5}(a), this model contains two narrow bands near the Fermi energy with similar feature as the TBG, providing a realistic platform for testing whether the mechanism identified in the Lieb-lattice analysis survives beyond the toy-model level.

The central result is that the anomaly persists in this realistic TBG calculation. As the chemical potential is tuned to the bottom of the narrow band, the normal-state conductance is strongly suppressed, while the critical current remains finite and can even increase as the conductance decreases. This is precisely the unconventional trend that motivates the present work and is the most important result of the realistic calculation [see Fig.~\ref{Fig5}(b)]. The six-band TBG model shows that $G$ and $I_c$ are no longer proportional to each other as understood in the conventional theory of Josephson junction. The conductance is still dominated by the dispersive channels, whereas the supercurrent is supported by the QMJC of the narrow-band. The fact that this behavior appears in a realistic 2D TBG Josehson junction, rather than only in the modified Lieb model, substantially strengthens the claim that the observed anomaly is an intrinsic consequence of narrow bands with nontrivial quantum metric.

The junction-length dependence provides an additional mechanistic test of this interpretation. As shown in Fig.~\ref{Fig5}(c), the conductance exhibits oscillatory behavior as the junction length increases, consistent with transport that remains sensitive to conventional propagating channels. In contrast, the critical current decays as the junction length increases [see Fig.~\ref{Fig5}(c)]. This is especially important because it shows that the interpretation is not based only on the $\mu$-dependence of $G$ and $I_c$, but is also supported by the length dependence of the critical current. In conclusion, both the chemical-potential dependence and the length dependence indicate that the critical current anomaly in realistic TBG arises from the competition between the conventional contribution and the quantum metric contribution, which is consistent with the mechanism identified in the Lieb-lattice analysis.
\\
\\
\textbf{Discussion}\\
In conclusion, we have demonstrated that, for a narrow band weak link with a non-trivial quantum metric, there exist two distinct contributions to the transport of normal current and supercurrent. The first is the conventional contribution arising from band dispersion, while the second originates from the quantum metric. Conventionally, in the long-junction and clean-junction regime, the normal current is transported through the propagating waves in the weak link, while the Josephson current as a proximity effect is suppressed by an exponential decay factor $e^{-L/\xi_{v_\mathrm{F}}}$ \cite{de2018superconductivity,tafuri2019fundamentals}, where $\xi_{v_\mathrm{F}}=\hbar v_\mathrm{F}/2\pi k_\mathrm{B}T$ is determined by the Fermi velocity $v_\mathrm{F}$. In contrast, both the conductance and Josephson current contributions associated with the quantum metric decay as $e^{-L/\xi_{\mathrm{loc}}}$, where $\xi_\mathrm{loc}$ is lower-bounded by the quantum metric length of the narrow band \cite{PhysRevResearch.7.023273}. When $\xi_\mathrm{loc}\gg\xi_{v_\mathrm{F}}$, normal current transport remains governed by band dispersion, while the Josephson current is dominated by the QMJC. Therefore, a decrease in conductance no longer indicates a decrease in critical Josephson current, due to the fundamentally different transport mechanisms, which constitutes the critical current anomaly proposed in this work.

Importantly, the conditions required for observing the critical current anomaly, namely a narrow band with a nontrivial quantum metric and small band dispersion, can be realized in Moir\'e materials such as twisted bilayer graphene. The recent experiment \cite{ccb4-tqxq} reports a relationship between $\mathcal{G}$ and $I_\mathrm{c}$ that closely resembles the anomaly proposed in this work.
\\
\\
\textbf{Methods}\\
For the realistic two-dimensional calculation, we model the weak link using the faithful six-band tight-binding Hamiltonian of twisted bilayer graphene as shown in Eq.~\ref{eq: 6-band TBG}. The superconducting leads are characterized by pairing gap $\Delta_{\mathrm{sc}} = 1\,\mathrm{meV}$, hopping amplitude $\lambda = 10\,\mathrm{meV}$, and lead chemical potential $\mu_{\mathrm{Lead}} = 0$. The lead--weak-link coupling is taken to be uniform over all orbitals at the interface with amplitude $10\,\mathrm{meV}$, and the induced gap between the narrow bands is $1.9\,\mathrm{meV}$. Unless otherwise stated, the junction size is $5 \times 30$ unit cells in the transverse and longitudinal directions, respectively, and the temperature is $k_{\mathrm{B}} T = 0.15\,\mathrm{meV}$. We compute the normal-state conductance $G$, the critical current $I_c$, and the length dependence of both quantities; the decay length of $I_c$ is extracted from a linear fit to the exponential length dependence. Technical details of the numerical implementation are provided in the Supplemental Material~\cite{supp}.

\section{Acknowledgment} K. T. L. acknowledge the support of the Ministry of Science and Technology, China, The New Cornerstone Foundation, the State Key Laboratory of Quantum Information Technologies and Materials, and the Hong Kong Research Grants Council through Grants No. MOST23SC01-A, No. RFS2021-6S03, No. C6053-23G, No. AoE/P-701/20, AoE/P-604/25R, No. 16309223, No. 16311424, and No. 16300325.

\bibliography{ref}

@article{hu2025anomalous,
  title={Anomalous coherence length in superconductors with quantum metric},
  author={Hu, Jin-Xin and Chen, Shuai A and Law, Kam Tuen},
  journal={Communications Physics},
  volume={8},
  number={1},
  pages={20},
  year={2025},
  publisher={Nature Publishing Group UK London},
  doi={https://doi.org/10.1038/s42005-024-01930-0}
}

@book{girvin2019modern,
  title={Modern condensed matter physics},
  author={Girvin, Steven M and Yang, Kun},
  year={2019},
  publisher={Cambridge University Press},
  address={Cornwall, Padstow, UK}
}

@article{l3c7-knqm,
  title = {Majorana Zero Modes in the Lieb-Kitaev Model with Tunable Quantum Metric},
  author = {Guo, Xingyao and Ma, Xinglei and Ying, Xuzhe and Law, K. T.},
  journal = {Phys. Rev. Lett.},
  volume = {135},
  issue = {7},
  pages = {076601},
  numpages = {6},
  year = {2025},
  month = {Aug},
  publisher = {American Physical Society},
  doi = {10.1103/l3c7-knqm},
  url = {https://link.aps.org/doi/10.1103/l3c7-knqm}
}

@article{PhysRevResearch.7.023273,
  title = {Flat band Josephson junctions with quantum metric},
  author = {Li, Zhong C. F. and Deng, Yuxuan and Chen, Shuai A. and Efetov, Dmitri K. and Law, K. T.},
  journal = {Phys. Rev. Res.},
  volume = {7},
  issue = {2},
  pages = {023273},
  numpages = {7},
  year = {2025},
  month = {Jun},
  publisher = {American Physical Society},
  doi = {10.1103/PhysRevResearch.7.023273},
  url = {https://link.aps.org/doi/10.1103/PhysRevResearch.7.023273}
}

@article{jiang2025superfluid,
  title={Superfluid weight cross-over and critical temperature enhancement in singular flat bands},
  author={Jiang, Guodong and T{\"o}rm{\"a}, P{\"a}ivi and Barlas, Yafis},
  journal={Proceedings of the National Academy of Sciences},
  volume={122},
  number={7},
  pages={e2416726122},
  year={2025},
  publisher={National Academy of Sciences},
  url={https://doi.org/10.1073/pnas.2416726122}
}

@ARTICLE{2023arXiv231015558D,
       author = {{Daido}, Akito and {Kitamura}, Taisei and {Yanase}, Youichi},
        title = "{Quantum geometry encoded to pair potentials}",
      journal = {arXiv e-prints},
         year = 2023,
        month = oct,
          eid = {arXiv:2310.15558},
        pages = {arXiv:2310.15558},
          doi = {10.48550/arXiv.2310.15558},
archivePrefix = {arXiv},
       eprint = {2310.15558},
 primaryClass = {cond-mat.supr-con},
       adsurl = {https://ui.adsabs.harvard.edu/abs/2023arXiv231015558D}
}

@article{PhysRevB.106.184507,
  title = {Quantum geometric effect on Fulde-Ferrell-Larkin-Ovchinnikov superconductivity},
  author = {Kitamura, Taisei and Daido, Akito and Yanase, Youichi},
  journal = {Phys. Rev. B},
  volume = {106},
  issue = {18},
  pages = {184507},
  numpages = {19},
  year = {2022},
  month = {Nov},
  publisher = {American Physical Society},
  doi = {10.1103/PhysRevB.106.184507},
  url = {https://link.aps.org/doi/10.1103/PhysRevB.106.184507}
}

@book{Datta_1995, place={Cambridge}, series={Cambridge Studies in Semiconductor Physics and Microelectronic Engineering}, title={Electronic Transport in Mesoscopic Systems}, publisher={Cambridge University Press}, author={Datta, Supriyo}, year={1995},address={New York}, collection={Cambridge Studies in Semiconductor Physics and Microelectronic Engineering}}

@article{PhysRevLett.66.3056,
  title = {Josephson current through a superconducting quantum point contact shorter than the coherence length},
  author = {Beenakker, C. W. J. and van Houten, H.},
  journal = {Phys. Rev. Lett.},
  volume = {66},
  issue = {23},
  pages = {3056--3059},
  numpages = {0},
  year = {1991},
  month = {Jun},
  publisher = {American Physical Society},
  doi = {10.1103/PhysRevLett.66.3056},
  url = {https://link.aps.org/doi/10.1103/PhysRevLett.66.3056}
}

@InProceedings{Beenakker_JJ,
author="Beenakker, C. W. J.",
editor="Fukuyama, Hidetoshi
and Ando, Tsuneya",
title="Three ``Universal'' Mesoscopic Josephson Effects",
booktitle="Transport Phenomena in Mesoscopic Systems",
year="1992",
publisher="Springer",
isbn="978-3-642-84818-6",
address={ Berlin, Heidelberg}
}

@article{peotta2015superfluidity,
       author = {{Peotta}, Sebastiano and {T{\"o}rm{\"a}}, P{\"a}ivi},
        title = "{Superfluidity in topologically nontrivial flat bands}",
      journal = {Nature Communications},
         year = 2015,
        month = nov,
       volume = {6},
          eid = {8944},
        pages = {8944},
          doi = {10.1038/ncomms9944}


}

@article{PhysRevLett.117.045303,
  title = {Geometric Origin of Superfluidity in the Lieb-Lattice Flat Band},
  author = {Julku, Aleksi and Peotta, Sebastiano and Vanhala, Tuomas I. and Kim, Dong-Hee and T\"orm\"a, P\"aivi},
  journal = {Phys. Rev. Lett.},
  volume = {117},
  issue = {4},
  pages = {045303},
  numpages = {6},
  year = {2016},
  month = {Jul},
  publisher = {American Physical Society},
  doi = {10.1103/PhysRevLett.117.045303},
  url = {https://link.aps.org/doi/10.1103/PhysRevLett.117.045303}
}

@article{PhysRevLett.131.240001,
  title = {Essay: Where Can Quantum Geometry Lead Us?},
  author = {T\"orm\"a, P\"aivi},
  journal = {Phys. Rev. Lett.},
  volume = {131},
  issue = {24},
  pages = {240001},
  numpages = {7},
  year = {2023},
  month = {Dec},
  publisher = {American Physical Society},
  doi = {10.1103/PhysRevLett.131.240001},
  url = {https://link.aps.org/doi/10.1103/PhysRevLett.131.240001}
}

@article{torma2022superconductivity,
       author = {{T{\"o}rm{\"a}}, P{\"a}ivi and {Peotta}, Sebastiano and {Bernevig}, Bogdan A.},
        title = "{Superconductivity, superfluidity and quantum geometry in twisted multilayer systems}",
      journal = {Nature Reviews Physics},
         year = 2022,
        month = jun,
       volume = {4},
       number = {8},
        pages = {528-542},
          doi = {10.1038/s42254-022-00466-y}
}

@article{PhysRevLett.132.026002,
  title = {Ginzburg-Landau Theory of Flat-Band Superconductors with Quantum Metric},
  author = {Chen, Shuai A. and Law, K. T.},
  journal = {Phys. Rev. Lett.},
  volume = {132},
  issue = {2},
  pages = {026002},
  numpages = {7},
  year = {2024},
  month = {Jan},
  publisher = {American Physical Society},
  doi = {10.1103/PhysRevLett.132.026002},
  url = {https://link.aps.org/doi/10.1103/PhysRevLett.132.026002}
}

@article{PhysRevB.103.144519,
  title = {Flat-band transport and Josephson effect through a finite-size sawtooth lattice},
  author = {Pyykk\"onen, Ville A. J. and Peotta, Sebastiano and Fabritius, Philipp and Mohan, Jeffrey and Esslinger, Tilman and T\"orm\"a, P\"aivi},
  journal = {Phys. Rev. B},
  volume = {103},
  issue = {14},
  pages = {144519},
  numpages = {12},
  year = {2021},
  month = {Apr},
  publisher = {American Physical Society},
  doi = {10.1103/PhysRevB.103.144519},
  url = {https://link.aps.org/doi/10.1103/PhysRevB.103.144519}
}

@article{balents2020superconductivity,
       author = {{Balents}, Leon and {Dean}, Cory R. and {Efetov}, Dmitri K. and {Young}, Andrea F.},
        title = "{Superconductivity and strong correlations in moir{\'e} flat bands}",
      journal = {Nature Physics},
         year = {2020},
        month = {May},
       volume = {16},
       number = {7},
        pages = {725-733},
          doi = {10.1038/s41567-020-0906-9}
}

@article{cao2018unconventional,
       author = {{Cao}, Yuan and {Fatemi}, Valla and {Fang}, Shiang and {Watanabe}, Kenji and {Taniguchi}, Takashi and {Kaxiras}, Efthimios and {Jarillo-Herrero}, Pablo},
        title = "{Unconventional superconductivity in magic-angle graphene superlattices}",
      journal = {\nat},
         year = 2018,
        month = apr,
       volume = {556},
       number = {7699},
        pages = {43-50},
          doi = {10.1038/nature26160}
}

@article{bistritzer2011moire,
       author = {{Bistritzer}, Rafi and {MacDonald}, Allan H.},
        title = "{Moir{\'e} bands in twisted double-layer graphene}",
      journal = {Proceedings of the National Academy of Science},
         year = 2011,
        month = jul,
       volume = {108},
       number = {30},
        pages = {12233-12237},
          doi = {10.1073/pnas.1108174108}
}

@article{park2021tunable,
       author = {{Park}, Jeong Min and {Cao}, Yuan and {Watanabe}, Kenji and {Taniguchi}, Takashi and {Jarillo-Herrero}, Pablo},
        title = "{Tunable strongly coupled superconductivity in magic-angle twisted trilayer graphene}",
      journal = {\nat},
         year = 2021,
        month = feb,
       volume = {590},
       number = {7845},
        pages = {249-255},
          doi = {10.1038/s41586-021-03192-0}
}

@article{torma2018quantum,
       author = {{T{\"o}rm{\"a}}, P. and {Liang}, L. and {Peotta}, S.},
        title = "{Quantum metric and effective mass of a two-body bound state in a flat band}",
      journal = {\prb},
         year = 2018,
        month = dec,
       volume = {98},
       number = {22},
          eid = {220511},
        pages = {220511},
          doi = {10.1103/PhysRevB.98.220511}
}

@article{provost1980riemannian,
       author = {{Provost}, J.~P. and {Vallee}, G.},
        title = "{Riemannian structure on manifolds of quantum states}",
      journal = {Communications in Mathematical Physics},
         year = 1980,
        month = sep,
       volume = {76},
       number = {3},
        pages = {289-301},
          doi = {10.1007/BF02193559}
}

@article{xie2020topology,
       author = {{Xie}, Fang and {Song}, Zhida and {Lian}, Biao and {Bernevig}, B. Andrei},
        title = "{Topology-Bounded Superfluid Weight in Twisted Bilayer Graphene}",
      journal = {\prl},
         year = 2020,
        month = apr,
       volume = {124},
       number = {16},
          eid = {167002},
        pages = {167002},
          doi = {10.1103/PhysRevLett.124.167002}
}

@article{PhysRevX.15.021056,
  title = {Topologically Protected Flatness in Chiral Moir\'e Heterostructures},
  author = {Cr\'epel, Valentin and Ding, Peize and Verma, Nishchhal and Regnault, Nicolas and Queiroz, Raquel},
  journal = {Phys. Rev. X},
  volume = {15},
  issue = {2},
  pages = {021056},
  numpages = {18},
  year = {2025},
  month = {May},
  publisher = {American Physical Society},
  doi = {10.1103/PhysRevX.15.021056},
  url = {https://link.aps.org/doi/10.1103/PhysRevX.15.021056}
}

@article{PhysRevB.98.134513,
  title = {Preformed pairs in flat Bloch bands},
  author = {Tovmasyan, Murad and Peotta, Sebastiano and Liang, Long and T\"orm\"a, P\"aivi and Huber, Sebastian D.},
  journal = {Phys. Rev. B},
  volume = {98},
  issue = {13},
  pages = {134513},
  numpages = {22},
  year = {2018},
  month = {Oct},
  publisher = {American Physical Society},
  doi = {10.1103/PhysRevB.98.134513},
  url = {https://link.aps.org/doi/10.1103/PhysRevB.98.134513}
}

@article{PhysRevB.101.060505,
  title = {Superfluid weight and Berezinskii-Kosterlitz-Thouless transition temperature of twisted bilayer graphene},
  author = {Julku, A. and Peltonen, T. J. and Liang, L. and Heikkil\"a, T. T. and T\"orm\"a, P.},
  journal = {Phys. Rev. B},
  volume = {101},
  issue = {6},
  pages = {060505},
  numpages = {7},
  year = {2020},
  month = {Feb},
  publisher = {American Physical Society},
  doi = {10.1103/PhysRevB.101.060505},
  url = {https://link.aps.org/doi/10.1103/PhysRevB.101.060505}
}

@article{verma2021optical,
  title={Optical spectral weight, phase stiffness, and ${T_c}$ bounds for trivial and topological flat band superconductors},
  author={Verma, Nishchhal and Hazra, Tamaghna and Randeria, Mohit},
  journal={Proceedings of the National Academy of Sciences},
  volume={118},
  number={34},
  pages={e2106744118},
  year={2021},
  publisher={National Acad Sciences},
  doi={10.1073/pnas.2106744118},
  url={https://doi.org/10.1073/pnas.2106744118}
}

@article{PhysRevB.106.014518,
  title = {Revisiting flat band superconductivity: Dependence on minimal quantum metric and band touchings},
  author = {Huhtinen, Kukka-Emilia and Herzog-Arbeitman, Jonah and Chew, Aaron and Bernevig, Bogdan A. and T\"orm\"a, P\"aivi},
  journal = {Phys. Rev. B},
  volume = {106},
  issue = {1},
  pages = {014518},
  numpages = {23},
  year = {2022},
  month = {Jul},
  publisher = {American Physical Society},
  doi = {10.1103/PhysRevB.106.014518},
  url = {https://link.aps.org/doi/10.1103/PhysRevB.106.014518}
}

@article{penttila2025flat,
  title={Flat-band ratio and quantum metric in the superconductivity of modified Lieb lattices},
  author={Penttil{\"a}, Reko PS and Huhtinen, Kukka-Emilia and T{\"o}rm{\"a}, P{\"a}ivi},
  journal={Communications Physics},
  volume={8},
  number={1},
  pages={50},
  year={2025},
  publisher={Nature Publishing Group UK London},
  url={https://doi.org/10.1038/s42005-025-01964-y}
}

@article{supp,
    author = {},
    title = {},
    year = {},
    volume = {},
    number = {},
    pages = {},
    journal = {See Supplemental Material for: (\romannumeral 1) Lattice Green's Function of Lieb Lattice, (\romannumeral 2) The Bound States in an Isolated Flat Band, (\romannumeral 3) Conductance and Critical Current of Isolated Flat Band with Nontrivial Quantum Metric, (\romannumeral 4) Details of the six-band TBG model},
    url={}
}

@article{PhysRevLett.130.216003,
  title = {Suppression of Nonequilibrium Quasiparticle Transport in Flat-Band Superconductors},
  author = {Pyykk\"onen, Ville A. J. and Peotta, Sebastiano and T\"orm\"a, P\"aivi},
  journal = {Phys. Rev. Lett.},
  volume = {130},
  issue = {21},
  pages = {216003},
  numpages = {7},
  year = {2023},
  month = {May},
  publisher = {American Physical Society},
  doi = {10.1103/PhysRevLett.130.216003},
  url = {https://link.aps.org/doi/10.1103/PhysRevLett.130.216003}
}

@article{ccb4-tqxq,
  title = {Probing the Flat-Band Limit of the Superconducting Proximity Effect in Twisted Bilayer Graphene Josephson Junctions},
  author = {D\'{\i}ez-Carl\'on, A. and D\'{\i}ez-M\'erida, J. and Rout, P. and Sedov, D. and Virtanen, P. and Banerjee, S. and Penttil\"a, R. P. S. and Altpeter, P. and Watanabe, K. and Taniguchi, T. and Yang, S.-Y. and Law, K. T. and Heikkil\"a, T. T. and T\"orm\"a, P. and Scheurer, M. S. and Efetov, D. K.},
  journal = {Phys. Rev. X},
  volume = {15},
  issue = {4},
  pages = {041033},
  numpages = {11},
  year = {2025},
  month = {Nov},
  publisher = {American Physical Society},
  doi = {10.1103/ccb4-tqxq},
  url = {https://link.aps.org/doi/10.1103/ccb4-tqxq}
}

@book{tafuri2019fundamentals,
  title={Fundamentals and Frontiers of the Josephson Effect},
  author={Tafuri, Francesco},
  volume={286},
  year={2019},
  publisher={Springer Nature},
  address={Cham, Switzerland}
}

@article{PhysRevLett.132.036001,
  title = {Spin-Triplet Superconductivity from Quantum-Geometry-Induced Ferromagnetic Fluctuation},
  author = {Kitamura, Taisei and Daido, Akito and Yanase, Youichi},
  journal = {Phys. Rev. Lett.},
  volume = {132},
  issue = {3},
  pages = {036001},
  numpages = {6},
  year = {2024},
  month = {Jan},
  publisher = {American Physical Society},
  doi = {10.1103/PhysRevLett.132.036001},
  url = {https://link.aps.org/doi/10.1103/PhysRevLett.132.036001}
}

@article{crepel2024chiral,
  title={Chiral limit and origin of topological flat bands in twisted transition metal dichalcogenide homobilayers},
  author={Cr{\'e}pel, Valentin and Regnault, Nicolas and Queiroz, Raquel},
  journal={Communications Physics},
  volume={7},
  number={1},
  pages={146},
  year={2024},
  publisher={Nature Publishing Group UK London},
url={https://doi.org/10.1038/s42005-024-01641-6}
}

@article{zhang2023visualizing,
  title={Visualizing moir{\'e} ferroelectricity via plasmons and nano-photocurrent in graphene/twisted-WSe2 structures},
  author={Zhang, Shuai and Liu, Yang and Sun, Zhiyuan and Chen, Xinzhong and Li, Baichang and Moore, SL and Liu, Song and Wang, Zhiying and Rossi, SE and Jing, Ran and others},
  journal={Nature communications},
  volume={14},
  number={1},
  pages={6200},
  year={2023},
  publisher={Nature Publishing Group UK London},
  url={https://doi.org/10.1038/s41467-023-41773-x}
}

@misc{yu2025quantumgeometryquantummaterials,
      title={Quantum Geometry in Quantum Materials}, 
      author={Jiabin Yu and B. Andrei Bernevig and Raquel Queiroz and Enrico Rossi and Päivi Törmä and Bohm-Jung Yang},
      year={2025},
      eprint={2501.00098},
      archivePrefix={arXiv},
      primaryClass={cond-mat.mes-hall},
      url={https://arxiv.org/abs/2501.00098}, 
}

@article{PhysRevLett.132.196202,
  title = {Strong-Coupling Phases of Trions and Excitons in Electron-Hole Bilayers at Commensurate Densities},
  author = {Dai, David D. and Fu, Liang},
  journal = {Phys. Rev. Lett.},
  volume = {132},
  issue = {19},
  pages = {196202},
  numpages = {6},
  year = {2024},
  month = {May},
  publisher = {American Physical Society},
  doi = {10.1103/PhysRevLett.132.196202},
  url = {https://link.aps.org/doi/10.1103/PhysRevLett.132.196202}
}

@article{park2025ferromagnetism,
  title={Ferromagnetism and topology of the higher flat band in a fractional Chern insulator},
  author={Park, Heonjoon and Cai, Jiaqi and Anderson, Eric and Zhang, Xiao-Wei and Liu, Xiaoyu and Holtzmann, William and Li, Weijie and Wang, Chong and Hu, Chaowei and Zhao, Yuzhou and others},
  journal={Nature Physics},
  pages={1--7},
  year={2025},
  publisher={Nature Publishing Group UK London},
  url={https://doi.org/10.1038/s41567-025-02804-0}
}

@article{PhysRevB.99.195455,
  title = {Faithful tight-binding models and fragile topology of magic-angle bilayer graphene},
  author = {Po, Hoi Chun and Zou, Liujun and Senthil, T. and Vishwanath, Ashvin},
  journal = {Phys. Rev. B},
  volume = {99},
  issue = {19},
  pages = {195455},
  numpages = {16},
  year = {2019},
  month = {May},
  publisher = {American Physical Society},
  doi = {10.1103/PhysRevB.99.195455},
  url = {https://link.aps.org/doi/10.1103/PhysRevB.99.195455}
}

@article{rossi2021quantum,
  title={Quantum metric and correlated states in two-dimensional systems},
  author={Rossi, Enrico},
  journal={Current Opinion in Solid State and Materials Science},
  volume={25},
  number={5},
  pages={100952},
  year={2021},
  publisher={Elsevier},
  url={https://doi.org/10.1016/j.cossms.2021.100952}
}

@article{PhysRevLett.124.076801,
  title = {Strange Metal in Magic-Angle Graphene with near Planckian Dissipation},
  author = {Cao, Yuan and Chowdhury, Debanjan and Rodan-Legrain, Daniel and Rubies-Bigorda, Oriol and Watanabe, Kenji and Taniguchi, Takashi and Senthil, T. and Jarillo-Herrero, Pablo},
  journal = {Phys. Rev. Lett.},
  volume = {124},
  issue = {7},
  pages = {076801},
  numpages = {7},
  year = {2020},
  month = {Feb},
  publisher = {American Physical Society},
  doi = {10.1103/PhysRevLett.124.076801},
  url = {https://link.aps.org/doi/10.1103/PhysRevLett.124.076801}
}

@ARTICLE{2022PhRvL.128h7002H,
       author = {{Herzog-Arbeitman}, Jonah and {Peri}, Valerio and {Schindler}, Frank and {Huber}, Sebastian D. and {Bernevig}, B. Andrei},
        title = "{Superfluid Weight Bounds from Symmetry and Quantum Geometry in Flat Bands}",
      journal = {\prl},
         year = 2022,
        month = feb,
       volume = {128},
       number = {8},
          eid = {087002},
        pages = {087002},
          doi = {10.1103/PhysRevLett.128.087002}
}

@ARTICLE{2018Natur.556...80C,
       author = {{Cao}, Yuan and {Fatemi}, Valla and {Demir}, Ahmet and {Fang}, Shiang and {Tomarken}, Spencer L. and {Luo}, Jason Y. and {Sanchez-Yamagishi}, Javier D. and {Watanabe}, Kenji and {Taniguchi}, Takashi and {Kaxiras}, Efthimios and {Ashoori}, Ray C. and {Jarillo-Herrero}, Pablo},
        title = "{Correlated insulator behaviour at half-filling in magic-angle graphene superlattices}",
      journal = {\nat},
         year = 2018,
        month = apr,
       volume = {556},
       number = {7699},
        pages = {80-84},
          doi = {10.1038/nature26154}
}

@ARTICLE{2020NatMa..19.1265A,
       author = {{Andrei}, Eva Y. and {MacDonald}, Allan H.},
        title = "{Graphene bilayers with a twist}",
      journal = {Nature Materials},
         year = 2020,
        month = nov,
       volume = {19},
       number = {12},
        pages = {1265-1275},
          doi = {10.1038/s41563-020-00840-0}
}

@ARTICLE{2021NatRM...6..201A,
       author = {{Andrei}, Eva Y. and {Efetov}, Dmitri K. and {Jarillo-Herrero}, Pablo and {MacDonald}, Allan H. and {Mak}, Kin Fai and {Senthil}, T. and {Tutuc}, Emanuel and {Yazdani}, Ali and {Young}, Andrea F.},
        title = "{The marvels of moir{\'e} materials}",
      journal = {Nature Reviews Materials},
         year = 2021,
        month = jan,
       volume = {6},
       number = {3},
        pages = {201-206},
          doi = {10.1038/s41578-021-00284-1}
}

@misc{verma2025quantumgeometryrevisitingelectronic,
      title={Quantum Geometry: Revisiting electronic scales in quantum matter}, 
      author={Nishchhal Verma and Philip J. W. Moll and Tobias Holder and Raquel Queiroz},
      year={2025},
      eprint={2504.07173},
      archivePrefix={arXiv},
      primaryClass={cond-mat.mtrl-sci},
      url={https://arxiv.org/abs/2504.07173}, 
}

@misc{dong2025flatbandexcitonsthreedimensional,
      title={Flat band excitons in a three-dimensional supertwisted spiral transition metal dichalcogenide}, 
      author={Yinan Dong and Yuzhou Zhao and Lennart Klebl and Taketo Handa and Ding Xu and Chiara Trovatello and Chennan He and Dihao Sun and Thomas P. Darlington and Kevin W. C. Kwock and Jakhangirkhodja A. Tulyagankhodjaev and Yusong Bai and Yinming Shao and Matthew Fu and Raquel Queiroz and Milan Delor and P. James Schuck and Xiaoyang Zhu and Tim O. Wehling and Song Jin and Eugene J. Mele and Dmitri N. Basov},
      year={2025},
      eprint={2506.21978},
      archivePrefix={arXiv},
      primaryClass={physics.app-ph},
      url={https://arxiv.org/abs/2506.21978}, 
}

@article{zondiner2020cascade,
  title={Cascade of phase transitions and Dirac revivals in magic-angle graphene},
  author={Zondiner, Uri and Rozen, Asaf and Rodan-Legrain, Daniel and Cao, Yuan and Queiroz, Raquel and Taniguchi, Takashi and Watanabe, Kenji and Oreg, Yuval and von Oppen, Felix and Stern, Ady and others},
  journal={Nature},
  volume={582},
  number={7811},
  pages={203--208},
  year={2020},
  publisher={Nature Publishing Group UK London},
  url={https://doi.org/10.1038/s41586-020-2373-y}
}

@ARTICLE{2021PNAS..11806744V,
       author = {{Verma}, Nishchhal and {Hazra}, Tamaghna and {Randeria}, Mohit},
        title = "{Optical spectral weight, phase stiffness, and T$_{c}$ bounds for trivial and topological flat band superconductors}",
      journal = {Proceedings of the National Academy of Science},
         year = 2021,
        month = aug,
       volume = {118},
       number = {34},
          eid = {e2106744118},
        pages = {e2106744118},
          doi = {10.1073/pnas.2106744118}
}

@ARTICLE{2021PhRvL.126b7002P,
       author = {{Peri}, Valerio and {Song}, Zhi-Da and {Bernevig}, B. Andrei and {Huber}, Sebastian D.},
        title = "{Fragile Topology and Flat-Band Superconductivity in the Strong-Coupling Regime}",
      journal = {\prl},
         year = 2021,
        month = jan,
       volume = {126},
       number = {2},
          eid = {027002},
        pages = {027002},
          doi = {10.1103/PhysRevLett.126.027002}
}

@ARTICLE{2019Natur.574..653L,
       author = {{Lu}, Xiaobo and {Stepanov}, Petr and {Yang}, Wei and {Xie}, Ming and {Aamir}, Mohammed Ali and {Das}, Ipsita and {Urgell}, Carles and {Watanabe}, Kenji and {Taniguchi}, Takashi and {Zhang}, Guangyu and {Bachtold}, Adrian and {MacDonald}, Allan H. and {Efetov}, Dmitri K.},
        title = "{Superconductors, orbital magnets and correlated states in magic-angle bilayer graphene}",
      journal = {\nat},
         year = 2019,
        month = oct,
       volume = {574},
       number = {7780},
        pages = {653-657},
          doi = {10.1038/s41586-019-1695-0}
}

@ARTICLE{2019Sci...363.1059Y,
       author = {{Yankowitz}, Matthew and {Chen}, Shaowen and {Polshyn}, Hryhoriy and {Zhang}, Yuxuan and {Watanabe}, K. and {Taniguchi}, T. and {Graf}, David and {Young}, Andrea F. and {Dean}, Cory R.},
        title = "{Tuning superconductivity in twisted bilayer graphene}",
      journal = {Science},
         year = 2019,
        month = mar,
       volume = {363},
       number = {6431},
        pages = {1059-1064},
          doi = {10.1126/science.aav1910}
}

@ARTICLE{2020NatPh..16..926S,
       author = {{Saito}, Yu and {Ge}, Jingyuan and {Watanabe}, Kenji and {Taniguchi}, Takashi and {Young}, Andrea F.},
        title = "{Independent superconductors and correlated insulators in twisted bilayer graphene}",
      journal = {Nature Physics},
         year = 2020,
        month = jun,
       volume = {16},
       number = {9},
        pages = {926-930},
          doi = {10.1038/s41567-020-0928-3}
}

@ARTICLE{2020Natur.583..375S,
       author = {{Stepanov}, Petr and {Das}, Ipsita and {Lu}, Xiaobo and {Fahimniya}, Ali and {Watanabe}, Kenji and {Taniguchi}, Takashi and {Koppens}, Frank H.~L. and {Lischner}, Johannes and {Levitov}, Leonid and {Efetov}, Dmitri K.},
        title = "{Untying the insulating and superconducting orders in magic-angle graphene}",
      journal = {\nat},
         year = 2020,
        month = jul,
       volume = {583},
       number = {7816},
        pages = {375-378},
          doi = {10.1038/s41586-020-2459-6}
}

@ARTICLE{2019PhRvL.123w7002H,
       author = {{Hu}, Xiang and {Hyart}, Timo and {Pikulin}, Dmitry I. and {Rossi}, Enrico},
        title = "{Geometric and Conventional Contribution to the Superfluid Weight in Twisted Bilayer Graphene}",
      journal = {\prl},
         year = 2019,
        month = dec,
       volume = {123},
       number = {23},
          eid = {237002},
        pages = {237002},
          doi = {10.1103/PhysRevLett.123.237002}
}

@ARTICLE{2019SciA....5.9770C,
       author = {{Codecido}, Emilio and {Wang}, Qiyue and {Koester}, Ryan and {Che}, Shi and {Tian}, Haidong and {Lv}, Rui and {Tran}, Son and {Watanabe}, Kenji and {Taniguchi}, Takashi and {Zhang}, Fan and {Bockrath}, Marc and {Lau}, Chun Ning},
        title = "{Correlated insulating and superconducting states in twisted bilayer graphene below the magic angle}",
      journal = {Science Advances},
         year = 2019,
        month = sep,
       volume = {5},
       number = {9},
        pages = {eaaw9770},
          doi = {10.1126/sciadv.aaw9770}
}

@ARTICLE{2019PhRvX...9c1049H,
       author = {{Hazra}, Tamaghna and {Verma}, Nishchhal and {Randeria}, Mohit},
        title = "{Bounds on the Superconducting Transition Temperature: Applications to Twisted Bilayer Graphene and Cold Atoms}",
      journal = {Physical Review X},
         year = 2019,
        month = jul,
       volume = {9},
       number = {3},
          eid = {031049},
        pages = {031049},
          doi = {10.1103/PhysRevX.9.031049}
}

@book{de2018superconductivity,
  title={Superconductivity of Metals and Alloys},
  author={De Gennes, Pierre-Gilles},
  year={2018},
  publisher={CRC press},
address={Boca Raton, Florida}
}

@article{PhysRevB.5.72,
  title = {Josephson Current Flow in Pure Superconducting-Normal-Superconducting Junctions},
  author = {Bardeen, John and Johnson, Jared L.},
  journal = {Phys. Rev. B},
  volume = {5},
  issue = {1},
  pages = {72--78},
  numpages = {0},
  year = {1972},
  month = {Jan},
  publisher = {American Physical Society},
  doi = {10.1103/PhysRevB.5.72},
  url = {https://link.aps.org/doi/10.1103/PhysRevB.5.72}
}
\end{document}